\documentclass[aps,prl,notitlepage,twocolumn,showpacs]{revtex4-1}

\usepackage{amsmath,amssymb}
\usepackage{graphicx}

\begin{document}

\newcommand{\bra}[1]    {\left\langle #1\right|}
\newcommand{\braket}[2]    {\left\langle #1\big|#2\right\rangle}
\newcommand{\ket}[1]    {| #1\rangle}
\newcommand{\av}[1]    {\big\langle #1 \big\rangle}
\newcommand{\tr}[1]     {{\rm Tr}\left[ #1 \right]}

\author{K. Gietka, P. Sza\'nkowski, T. Wasak and J. Chwede\'nczuk}
\affiliation{Faculty of Physics, University of Warsaw, ul. Pasteura 5, 02--093 Warsaw, Poland}

\title{Quantum-enhanced interferometry and the structure of twisted states}

\begin{abstract}
  Preparation of a non-classically correlated state is the first step of any quantum-enhanced interferometric protocol.
  An efficient method is the one-axis twisting, which entangles a collection of initially uncorrelated particles by means of two-body interactions.
  Here we investigate the limits of the quantum improvement which can be reached with this method in realistic experimental conditions.
  We demonstrate that the usefulness of this entangling mechanism is a result of fine structures introduced into the quantum state.
  The scale at which these structures vary allows us to identify the minimal requirements for the precision of the complete interferometric sequence.
%  In particular, we show that the quantum enhancement does not necessarily require the single-particle detection efficiency.
  Our results---especially the explanation of the underlying principle of the entangling method---may help to develop ultra-precise interferometers.
\end{abstract}


\maketitle

{\it Introduction}---Generation of many-body entangled states for future applications in precise interferometry is a challenging task.
With photons, it can be accomplished by creating a squeezed state of light which is
fed into one of the interferometer's arms \cite{knight_qo}. Here squeezing refers to the reduction of the  fluctuations of one of the quadratures of the electromagnetic field.
The concept of squeezing can be---to some extent---transferred to the atomic domain.
However, since coherences between different atom number states are forbidden \cite{ssr_wick}, a single-mode squeezed state of massive particles does not exist.
Nevertheless, squeezing---in this context called spin-squeezing---can be generated by reducing the fluctuations of the atom number difference between the two arms of the interferometer
\cite{kitagawa1993squeezed, wineland1994squeezed}.
This reduction is usually achieved using the repulsive two-body interactions between the atoms \cite{sorensen2001many}.

In the pioneering work \cite{esteve2008squeezing}, spin-squeezing was observed in a collection of ultra-cold rubidium atoms trapped in a double-well potential. 
The following experiments employed a technique called one-axis twisting (OAT) to squeeze the fluctuations between two hyper-fine states
of ultra-cold atoms \cite{appel2009mesoscopic,gross2010nonlinear,riedel2010atom,oat1,nori,polzik}. In this approach, first discussed in \cite{kitagawa1993squeezed},
particles in a coherent two-mode state \cite{radcliffe1971some} are dynamically entangled due to collisions. Spin-squeezing emerges immediately as the interactions start to operate.
At some moment, however, spin-squeezing is lost,
yet this does not mean the state is useless for precise interferometry. To contrary, the amount of useful entanglement continues to grow \cite{pezze2009entanglement},
reaching the maximal attainable value \cite{greenberger1989going}. 
This effect---loss of squeezing accompanied by the increase of the entanglement useful for interferometry witnessed by the Fisher information---has been recently observed in the experiment \cite{smerzi_ob}. 

Interferometry with matter-waves seems
particularly promising for future realistic metrological applications \cite{schmied_rmp}. This is because massive particles
can strongly interact with the external field imprinting the phase, giving high interferometric signal.
Moreover, two-body collisions of atoms allow for creation
of many-body entangled states without the mediating element, such as a crystal used for the parametric down conversion of light \cite{pdc1,pdc2}.

In this work, we identify the underlying mechanism responsible for high interferometric efficiency of the OAT states.
We consider two emblematic transformations---the Mach-Zehnder interferometer (MZI) and a generalized beam-splitter (BS)---both accessible with current experimental techniques
\cite{appel2009mesoscopic,gross2010nonlinear,riedel2010atom,smerzi_ob,berrada2013integrated,smerzi_twin}. We show that these interferometers,
combined with the measurement of the number of particles in each output port, provide an ultra-high phase sensitivity. We also discuss the influence of the detection imperfections.
Those come from the limited control over the interaction strength or the duration of the OAT procedure as well as the imprecise measurement of the number of particles. 
Our results contribute to the understanding of the general features of the quantum nonlinear rotator \cite{rotator}. They may also find application to the developing field of quantum-enhanced 
interferometry operating on cold-atom ensembles \cite{polzik, non_condensed} and Bose-Einstein condensates. The experimental achievement of \cite{smerzi_ob} underlines the practical relevance of 
the entangled---but not spin-squeezed---states. The content of our work emphasizes the practical significance of even further-correlated OAT states for ultra-precise interferometry. Other authors studied 
the role of decoherence in the OAT procedure (see for example \cite{kolo,ferrini}). We focus on the implementation of the OAT states and emphasize the role of the precise detection 
in maintaining high sensitivity.

{\it Twisted states and quantum interferometry}---Usually, an interferometer consists of two arms through which either a pulse of light- or a matter-wave moves. During this propagation, a relative
phase $\theta$ is imprinted between the arms which later intersect to give an interference signal. From this output, the value of $\theta$ is deduced. 

The OAT is a method of preparation of an entangled state, which is injected into an interferometer and it works as follows. 
Take a pure Bose-Einstein condensate in an equal superposition of modes $a$ and $b$, i.e.,
\begin{equation}\label{cohmix}
  \ket{\psi}=\frac1{\sqrt{N!}}\left(\frac{\hat a^\dagger+\hat{b}^\dagger}{\sqrt 2}\right)^N\ket{0},
\end{equation}
where $\hat a^\dagger$ and $\hat b^\dagger$ are the corresponding creation operators.
The goal is to transform this separable state of $N$ particles into an entangled one.
This is achieved in the absence of tunneling between the modes, when the evolution is triggered by the two-body interactions.
We denote by $\chi^{-1}$ the associated timescale and have the two-mode Hamiltonian
\begin{equation}\label{ham}
  \hat H=\hbar\chi\hat J_z^2.
\end{equation}
Here $\hat J_z$ is one member of a triad of angular momentum operators $\hat J_x=\frac12(\hat a^\dagger\hat b+\hat b^\dagger\hat a)$, $\hat J_y=\frac1{2i}(\hat a^\dagger\hat b-\hat b^\dagger\hat a)$, and
$\hat J_z=\frac12(\hat a^\dagger\hat a-\hat b^\dagger\hat b)$.
The evolution of the initial coherent state (\ref{cohmix}) with the Hamiltonian (\ref{ham}) gives
\begin{equation}\label{twist}
  \ket{\psi_\alpha}= e^{-i \alpha \hat J_z^2} \ket\psi,
\end{equation}
where  $\alpha = \chi t$. The effect of interactions can be pictured using the mode occupation states $\ket m$, which denote $\frac N2+m$ particles in mode $a$ and $\frac N2-m$ in $b$ with
$m\in[-\frac N2,\frac N2]$. It this basis, the state (\ref{twist}) reads
\begin{equation}\label{ketm}
  \ket{\psi_\alpha}=\sum_{m=-\frac N2}^{\frac N2}e^{-i\alpha m^2}C_m\ket m,
\end{equation}
with $C_m=\sqrt{\frac1{2^N}\binom N{\frac N2+m}}$.
The action of the Hamiltonian (\ref{ham}) imprints the $\alpha$-dependent phase which is quadratic in $m$. For most values of $\alpha$, this phase jumps rapidly between the states $\ket m$
and $\ket{m+1}$. This is an important property of the OAT, and it is useful for the analytical calculations.

\begin{figure}[t!]
  \includegraphics[clip, scale=0.55]{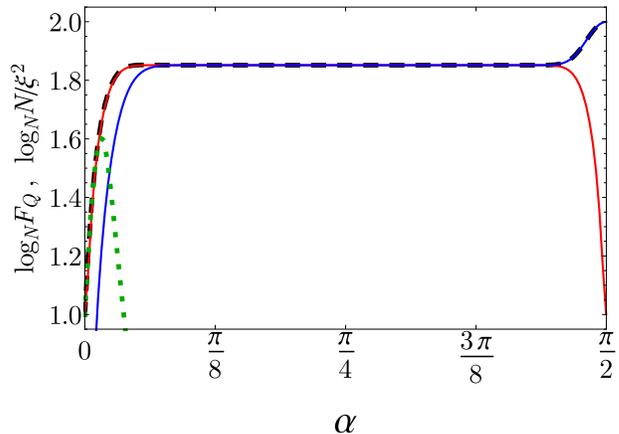}
  \caption{(color online) The QFI from Eq.~(\ref{pure}) optimized over the direction of $\vec n$ (dashed black line) compared to the QFI for the BS (solid blue line) and for the MZI
    (solid red line). The inverse of $\xi^2$ calculated using Eq.~(\ref{ss}) optimized over the direction of $\vec n_1$ is shown with a green dotted line. 
    All quantities are drawn for $N=100$ particles.
    \label{panel}}
\end{figure}

The OAT procedure transforms the coherent state (\ref{cohmix}) into the entangled one (\ref{twist}) and is usually considered in the relation to the spin-squeezing parameter
\cite{wineland1994squeezed,sorensen2001many}
\begin{equation}\label{ss}
  \xi^2=N\frac{\av{(\Delta \hat J_{n_1})^2}}{\langle \hat{J}_{n_2}\rangle^2+\langle \hat{J}_{n_3}\rangle^2}.
\end{equation}
Each $\hat J_{n_i}=\hat{\vec J}\cdot\vec n_i$ is a scalar product of the vector of angular momentum operators $\hat{\vec J}=[\hat J_x,\hat J_y,\hat J_z]$ and a unit vector $\vec n_i$
($\vec n_1$, $\vec n_2$, and $\vec n_3$ are orthogonal), while $\Delta \hat J_{n_1}=\hat J_{n_1}-\av{\hat J_{n_1}}$.
When $\xi^2<1$, the state is called spin-squeezed, and it follows that it is
particle-entangled and useful for sub shot-noise (SSN) metrology \cite{pezze2009entanglement,giovannetti2006quantum}. 
This is because the phase sensitivity can be linked to the spin-squeezing by the formula $\Delta\theta\propto\xi/{\sqrt N}$.
Note that $\xi^2<1$ can happen only when the denominator of Eq.~(\ref{ss}) is large. Since the angular momentum operators couple the neighbouring states $\ket m$ and $\ket{m+1}$, 
then, according to Eq.~(\ref{ketm}), their average value contains phase factors $e^{\pm2im\alpha}$. Apart from $\alpha\simeq k\,\pi$ with $k\in\mathbb{Z}$, 
these phase factors quickly oscillate. As a result, the denominator of Eq.~(\ref{ss}) is strongly suppressed. 
This means that within one period of evolution the state can be significantly spin-squeezed only for small $\alpha$ (i.e., short times),
as observed in the experiment \cite{smerzi_ob} and in agreement with the numerically calculated $\xi^2$,
which optimized over the direction of $\vec n_1$ at each $\alpha$ is drawn in Fig.~\ref{panel} with dotted green line. Analytical estimates show that indeed the OAT generates spin-squeezed states only for
$\alpha\lesssim N^{-\frac12}$ \cite{pezze2009entanglement}, when the phase variations are smooth on the $m\rightarrow m+1$ increment
\footnote{Indeed, the variation of the phase $\varphi(m)=\alpha m^2$ can be estimated by its derivative $\frac{\partial\varphi}{\partial m}\lesssim2\pi$. For the initial coherent state, the maximal 
$m\sim N^{\frac12}$, giving $\alpha\lesssim N^{-\frac12}$}. This means that for large $N$ the presence of spin-squeezing does not last long.

Although  $\xi^2>1$ at later times, the state (\ref{twist}) is still useful for SSN interferometry \cite{pezze2009entanglement,smerzi_twin}. To demonstrate this, note that an interferometer 
imprints $\theta$  through a unitary transformation
\begin{equation}\label{anyinterferometer}
  \hat U_n(\theta) = e^{-i \theta \hat{J}_{n} }.
\end{equation}
Similarly to Eq.~(\ref{ss}), $\hat J_{n}=\hat{\vec J}\cdot\vec n$ while different directions of $\vec n$ refer do distinct interferometric transformations.
When $\vec n$ is oriented along the $x$ axis, thus $\hat{J}_{n}=\hat J_x$, the mode mixing
transformation represents a generalized  BS, while for $\hat{J}_{n}=\hat J_y$, it stands for the MZI.
Finally, for $\hat{J}_{n}=\hat J_z$, the interferometer only imprints the phase $\theta$ between the modes without mixing them.
Any other direction of $\vec n$ can be expressed as a sequence of these transformations.

If $\theta$ is estimated through a series of $\mu\gg1$ experiments, the precision of estimation, according to the Cramer-Rao
Lower Bound \cite{holevo2011probabilistic}, cannot be better than
\begin{equation}\label{crlb}
  \Delta \theta= \frac1{\sqrt\mu}\frac1{\sqrt{ F_Q}}.
\end{equation}
Here, the $F_Q$ is the quantum Fisher information (QFI) \cite{braunstein1994statistical}, which tells how much information about $\theta$ can be encoded in a state $\hat\varrho$
through a transformation (\ref{anyinterferometer}). For a pure state, such as in Eq.~(\ref{twist}), it is given by four times the variance of the generator $\hat J_{n}$ calculated in this state,
\begin{equation}\label{pure}
  F_Q=4\av{(\Delta \hat J_{n})^2}.
\end{equation}
When $F_Q=N$,  the lower bound for the sensitivity is the shot-noise limit (SNL)  $\Delta\theta=\frac1{\sqrt\mu}\frac1{\sqrt N}$. According to Eq.~(\ref{crlb}), when the QFI is large, higher
sensitivity can be reached, but $F_Q>N$ is only possible---allowing for SSN sensitivity---when the initial state
is particle-entangled \cite{pezze2009entanglement}.
The QFI calculated for the state (\ref{twist}) is drawn with a dashed black line in Fig.~\ref{panel} for each $\alpha$ maximized over the direction $\vec n$.
It has a characteristic shape---it grows from the value of $F_Q=N$ for the initially uncorrelated state (\ref{cohmix}) to reach a wide plateau approximately at a moment when the spin-squeezing vanishes.
Finally, through a narrow peak it strictly reaches the Heisenberg limit
$F_Q=N^2$ at $\alpha=\frac\pi2$ and then symmetrically drops to $F_Q=N$ at $\alpha=\pi$ \cite{pezze2009entanglement}.
This plot shows that for the state (\ref{twist}), $F_Q\geqslant N$ for all $\alpha$.

To better understand why the QFI stays at a constant and very high level for most of the time, we focus on two types of interferometric transformations such that
can be implemented in the experiment, i.e., the BS and the MZI (rotations around the $x$ and $y$ axis, respectively). When $\alpha\gtrsim N^{\frac12}$---so that the phase in Eq.~(\ref{ketm})
quickly jumps between the neighboring states---we have \footnote{We have used the fact that the $C_m$'s and the imprinted phase are symmetric with respect to $m=0$.}
\begin{subequations}
  \begin{eqnarray}
    F_Q&=&4\av{(\Delta\hat J_{x/y})^2}\simeq2\Big[\sum_{m=-\frac N2}^{\frac N2}C_m^2\beta_{m}^2+\label{fast1}\\
      &\pm&\sum_{m=-\frac N2}^{\frac N2}C_mC_{m+2}\beta_{m}\beta_{m+1}e^{4i\alpha m}\Big],\label{fast}
  \end{eqnarray}
\end{subequations}
where $\beta_{m}=\sqrt{(N/2+m+1)(N/2-m)}$, which for large $N$ changes slowly as compared to $C_m$'s and can be approximated by $\frac N2$. Moreover, when $\alpha$ is not too close to $\frac\pi2$, the quickly
oscillating phase ``kills'' the line (\ref{fast}), giving $F_Q=\frac12N^2$ both for the BS and the MZI. This happens for most values of $\alpha$, which explains the presence of the wide plateau
in Fig.~\ref{panel}. Only when $\alpha\simeq\frac\pi2$, line (\ref{fast}) starts to contribute since the phase
factor is then $\sim e^{i2\pi m}=1$. For the BS, the contribution of the line (\ref{fast}) is positive, so it will increase the value of the QFI up to the Heisenberg limit at $\alpha=\frac\pi2$.
However, the MZI due to the opposite sign in line (\ref{fast})
will experience a major drop of the QFI. These analytical estimates fully agree with the numerical results shown in Fig.~\ref{panel}.

Equipped with this knowledge, we show that the estimation of $\theta$ from the measurement of the population imbalance between the two output ports of either the MZI or the BS interferometer
exploits the entanglement of the state (\ref{twist}). To this end, we calculate the related Fisher information (FI) \cite{holevo2011probabilistic} given by the formula
\begin{equation}\label{cfi}
  F(\theta)=\sum_{m=-\frac N2}^{\frac N2}\frac1{p(m|\theta)}\left(\frac{\partial p(m|\theta)}{\partial\theta}\right)^2,
\end{equation}
where the probability $p(m|\theta)$ is the projection of the output state of the interferometer onto the state $\ket{m}$, i.e.,
\begin{equation}\label{prob}
p(m|\theta)=|\!\bra{m}\hat U_n(\theta)\ket{\psi_\alpha}|^2.
\end{equation}
The FI measures the amount of information about $\theta$ contained in the probability $p(m|\theta)$, and the corresponding CRLB, in analogy to Eq.~(\ref{crlb}), reads
\begin{equation}\label{crlb2}
  \Delta \theta= \frac1{\sqrt \mu}\frac1{\sqrt{F(\theta)}}.
\end{equation}
Since the QFI sets the maximal attainable sensitivity for any measurement, $F(\theta)\leqslant F_Q$ holds \cite{braunstein1994statistical}.
For both BS and MZI transformations---chosen by aligning the direction of the interferometric rotation $\vec n$ by either the $x$ or $y$ axis in Eq.~(\ref{prob})
and plugging the resulting probability into the Eq.~(\ref{cfi})---we calculate
the FI as a function of $\theta$ for each $\alpha$. 
For instance, for the BS, the FI is
\begin{subequations}\label{fi}
\begin{eqnarray}
    F(\theta)&=&\sum_{m=-\frac N2}^{\frac N2}\Big(\beta_m\sin(\tilde\varphi_{m+1}-\tilde\varphi_m)\big|\tilde C_{m+1}\big|+\label{fi_a}\\
    &-&\beta_{m-1}\sin(\tilde\varphi_{m}-\tilde\varphi_{m-1})\big|\tilde C_{m-1}\big|\Big)^2.\label{fi_b}
\end{eqnarray}
\end{subequations}
In this expression, $\tilde C_{m}=\bra m\hat U_x(\theta)\ket{\psi_\alpha}$ and $\tilde\varphi_m$ is its phase. On the plateau, when this phase
quickly jumps between the neighbouring states, the contribution to the FI resulting from the multiplication of the line (\ref{fi_a}) by (\ref{fi_b})
will be negligible. Moreover, the squared sine functions can be approximated by its average value $1/2$, which leads to
\begin{equation}
    F(\theta)\simeq\sum_{m=-\frac N2}^{\frac N2}\beta_m^2\big|\tilde C_m\big|^2.
\end{equation}
This expression closely resembles Eq.~(\ref{fast1}) and differs only by a constant factor of two. When $\theta\simeq0$, so that $\tilde C_m$ remains peaked
around $m=0$, we can approximate $\beta_m^2\simeq N^2/4$ which gives 
\begin{equation}\label{fin2}
    F\simeq\frac{N^2}4\sum_{m=-\frac N2}^{\frac N2}\big|\tilde C_m\big|^2=\frac{N^2}4.
\end{equation}
Note that for the MZI, in Eq.~(\ref{fi}), the sine functions are replaced with
the cosine and the minus sign between lines (\ref{fi_a}) and (\ref{fi_b}) is replaced by a plus. This however has no impact on the above discussion giving the same
value of the FI as in Eq.~(\ref{fin2}). Therefore, for both interferometers, the estimation from the population imbalance is very efficient 
and gives the Heisenberg scaling for a wide range of $\alpha$'s. The outcome of our analytical considerations fully agrees with the numerical calculations of the FI directly from Eq.~(\ref{cfi}).

The result (\ref{fin2}) is obtained for the ideal case, in absence of detection imperfections.
We now make a final step and analyze the influence of such limitations.

{\it Influence of detection imperfections}---We consider two sources of imperfections which might have a dominating
impact on the performance of the OAT-based interferometer. The first one results from an inadequate control over the value of $\alpha$. This might be a consequence of the limited knowledge about
the strength of the two-body interactions (for instance resulting from the fluctuating magnetic field \cite{fesch_jul,dalfovo,zwerger,szan}) 
or the lack of precise control over the duration of the OAT procedure.
We account for this effect by assuming that $\alpha$, rather than fixed, fluctuates with a Gaussian probability with uncertainty $\delta\alpha$.
In consequence, we replace the pure state from Eq.~(\ref{twist}) with a mixture
\begin{equation}
    \hat\varrho_\alpha=\int\! d\tilde\alpha\, P(\tilde\alpha-\alpha)\ket{\psi_{\tilde\alpha}}\bra{\psi_{\tilde\alpha}},
\end{equation}
with $P(\tilde\alpha)\propto\exp\left[-\tilde\alpha^2/2(\delta\alpha)^2\right]$. The FI from Eq.~(\ref{cfi}) is now based on
the probability distribution $p(m|\theta)=\tr{\ket m\bra m \hat U_n(\theta)\hat\varrho_\alpha \hat U^\dagger_n(\theta)}$. It is reasonable to assume that
$\delta\alpha\lesssim1$, otherwise it would be not possible to resolve even the wide plateau of Fig.~\ref{panel}. In this limit, the result
of Eq.~(\ref{fin2}) is replaced by
\begin{equation}
    F\simeq\frac{N^2}4\times\frac1{\sqrt{1+2N(\delta\alpha)^2}}.
\end{equation}
If the control over $\alpha$ is high so that $\delta\alpha\lesssim N^{-\frac12}$, the scaling proportional to $N^2$ is retained. On the other hand, when
$\delta\alpha\gtrsim N^{-\frac12}$ (or even is of the order of unity), the scaling is $F\propto N^{\frac32}$, which still is still much better than the SNL.

\begin{figure}[t!]
  \includegraphics[clip, scale=0.55]{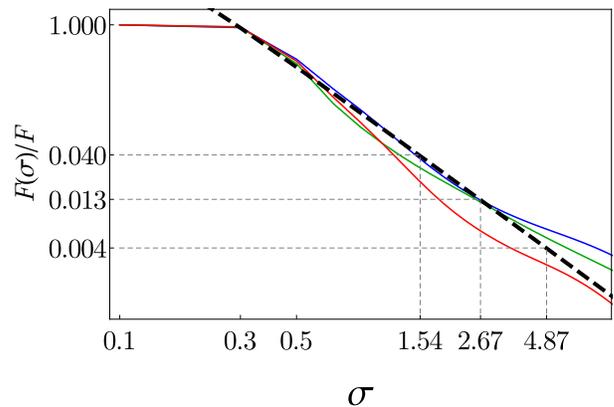}
  \caption{(color online) The FI with finite atom detection resolution $\sigma$, normalized to the FI with perfect accuracy. Three different solid lines correspond to different values of $\theta\simeq0$
    (i.e., $\theta=0.02\pi$, 0.07$\pi$ and $0.14\pi$). The dashed black line is the best fit giving a function $0.095\sigma^{-2}$. The horizontal dashed grey lines indicate the SNL for $N=100$, 
    300 and 1000. 
    \label{sigma}}
\end{figure}

Another factor which might reduce the sensitivity is the finite resolution of the atom detection $\sigma$ \cite{obert_detect,sheet,bloch,paris_han}.
To account for this, we replace the probability (\ref{prob}) with
\begin{equation}\label{reso}
    \tilde p(m|\theta)=\sum_{m'=-\frac N2}^{\frac N2}\!\!\!R(m-m')\,p(m'|\theta),
\end{equation}
where we take $R(m)\propto\exp[-m^2/2\sigma^2]$ as the resolution function. %% We substitute (\ref{reso}) into Eq.~(\ref{cfi}) and assuming $\sigma_r\gtrsim1$ we obtain at the plateau
%% \begin{equation}\label{res}
%%     F\simeq\frac{N^2}4e^{-2\alpha^2\sigma_r^2}.
%% \end{equation}
The finite resolution may have a much more profound effect on the FI than the uncertainty in determination of $\alpha$.
This is because on the plateau, its characteristic rapid phase variations summed even over a small increment of $m'$ oscillate down to zero.
This intuitive picture is confirmed by the numerically obtained $F(\sigma)$ calculated at the plateau for different values of $\theta$, see Fig.~\ref{sigma}.
We observe that the value of FI has a universal behavior
\begin{equation}
  F(\sigma)\propto\frac{N^2}4\times\frac1{\sigma^2}.
\end{equation}
This $\sigma$-dependent scaling function does not depend on $N$ because it is rather related to the fast variation of the phase on the $m'\rightarrow m'+1$ increment in Eq.~(\ref{reso}). For given $N$, this
expression allows for a simple estimate of a minimal required resolution to beat the SNL. In general, $\sigma\lesssim\sqrt N$ is necessary to retain the SSN sensitivity, 
see the horizontal dashed lines in Fig.~\ref{sigma}.

The above result is an example of a more general feature of entangled states. From the metrological point of view, such states are highly susceptible to the parameter-imprinting transformations \cite{wasak2015interferometry}. This property is witnessed by the fidelity \footnote{The fidelity between two probability distributions $p_n$ and $q_n$ is defined as $f=\sum_n\sqrt{p_nq_n}$.} between two neighboring probabilities $p(m|\theta)$ and $p(m|\theta+\delta\theta)$ \cite{smerzi_ob}. 
Generally speaking, probabilities which have fine structures are more prone to changes when $\theta$ is varied. 
In our case, those fine structures can be found in the rapid variations of the phase, which are
governed by the value of $\alpha$. It is now clear that on the plateau, when $\alpha$ is high, so that the phase changes abruptly, a sophisticated detection technique is required to grasp the quantumness 
of the state $\ket{\psi_\alpha}$.

{\it Conclusions}---We have shown that the rapid phase variations of the one-axis-twisted state is the underlying mechanism responsible for its high interferometric efficiency.
We have focused on two experimentally accessible transformations. The measurement of the population imbalance between the output ports provides a high sensitivity, which might 
be altered by the experimental imperfections. The lack of knowledge about the precise value of $\alpha$ is not as severe, as compared to the inaccurate atom number detection. 
The presented discussion, in particular the explanation of the essential principle of the one-axis-twisting method, might find application in other areas of ultra-precise metrology.

\begin{acknowledgments}{\it  Acknowledgments}---Discussions with Emilia Witkowska and Darek Kajtoch are ackonwledged. This work was supported by
the Polish Minsitry of Science and Higher Education programme ``Iuventus Plus'' (2015-2017), project number IP2014 050073 and
the National Science Center grant no. DEC-2011/03/D/ST2/00200.
\end{acknowledgments}


\begin{thebibliography}{10}%
\makeatletter
\providecommand \@ifxundefined [1]{%
 \ifx #1\undefined \expandafter \@firstoftwo
 \else \expandafter \@secondoftwo
\fi
}%
\providecommand \@ifnum [1]{%
 \ifnum #1\expandafter \@firstoftwo
 \else \expandafter \@secondoftwo
\fi
}%
\providecommand \enquote [1]{``#1''}%
\providecommand \bibnamefont  [1]{#1}%
\providecommand \bibfnamefont [1]{#1}%
\providecommand \citenamefont [1]{#1}%
\providecommand\href[0]{\@sanitize\@href}%
\providecommand\@href[1]{\endgroup\@@startlink{#1}\endgroup\@@href}%
\providecommand\@@href[1]{#1\@@endlink}%
\providecommand \@sanitize [0]{\begingroup\catcode`\&12\catcode`\#12\relax}%
\@ifxundefined \pdfoutput {\@firstoftwo}{%
 \@ifnum{\z@=\pdfoutput}{\@firstoftwo}{\@secondoftwo}%
}{%
 \providecommand\@@startlink[1]{\leavevmode}%
 \providecommand\@@endlink[0]{}%
}{%
 \providecommand\@@startlink[1]{%
  \leavevmode
  \pdfstartlink
   attr{/Border[0 0 1 ]/H/I/C[0 1 1]}%
   user{/Subtype/Link/A<</Type/Action/S/URI/URI(#1)>>}%
  \relax
 }%
 \providecommand\@@endlink[0]{\pdfendlink}%
}%
\providecommand \url  [0]{\begingroup\@sanitize \@url }%
\providecommand \@url [1]{\endgroup\@href {#1}{\urlprefix}}%
\providecommand \urlprefix [0]{URL }%
\providecommand \Eprint[0]{\href }%
\@ifxundefined \urlstyle {%
  \providecommand \doi [1]{doi:\discretionary{}{}{}#1}%
}{%
  \providecommand \doi [0]{doi:\discretionary{}{}{}\begingroup
  \urlstyle{rm}\Url }%
}%
\providecommand \doibase [0]{http://dx.doi.org/}%
\providecommand \Doi[1]{\href{\doibase#1}}%
\providecommand \bibAnnote [3]{%
  \BibitemShut{#1}%
  \begin{quotation}\noindent
    \textsc{Key:}\ #2\\\textsc{Annotation:}\ #3%
  \end{quotation}%
}%
\providecommand \bibAnnoteFile [2]{%
  \IfFileExists{#2}{\bibAnnote {#1} {#2} {\input{#2}}}{}%
}%
\providecommand \typeout [0]{\immediate \write \m@ne }%
\providecommand \selectlanguage [0]{\@gobble}%
\providecommand \bibinfo [0]{\@secondoftwo}%
\providecommand \bibfield [0]{\@secondoftwo}%
\providecommand \translation [1]{[#1]}%
\providecommand \BibitemOpen[0]{}%
\providecommand \bibitemStop [0]{}%
\providecommand \bibitemNoStop [0]{.\EOS\space}%
\providecommand \EOS [0]{\spacefactor3000\relax}%
\providecommand \BibitemShut [1]{\csname bibitem#1\endcsname}%
%</preamble>
\bibitem{knight_qo}%
  \BibitemOpen
  \bibfield{author}{%
  \bibinfo {author} {\bibfnamefont{C.}~\bibnamefont{Gerry}}\ and\ \bibinfo
  {author} {\bibfnamefont{P.}~\bibnamefont{Knight}},\ }%
  \emph{\bibinfo {title} {Introductory Quantum Optics}}\ (\bibinfo {publisher}
  {Cambridge University Press},\ \bibinfo {year} {2004})%
  \bibAnnoteFile{NoStop}{knight_qo}%
\bibitem{ssr_wick}%
  \BibitemOpen
  \bibfield{author}{%
  \bibinfo {author} {\bibfnamefont{G.~C.}\ \bibnamefont{Wick}}, \bibinfo
  {author} {\bibfnamefont{A.~S.}\ \bibnamefont{Wightman}},\ and\ \bibinfo
  {author} {\bibfnamefont{E.~P.}\ \bibnamefont{Wigner}},\ }%
  \bibfield{journal}{%
  \bibinfo {journal} {Phys. Rev.}\ }%
  \textbf{\bibinfo {volume} {88}},\ \bibinfo {pages} {101} (\bibinfo {year}
  {1952})%
  \bibAnnoteFile{NoStop}{ssr_wick}%
\bibitem{kitagawa1993squeezed}%
  \BibitemOpen
  \bibfield{author}{%
  \bibinfo {author} {\bibfnamefont{M.}~\bibnamefont{Kitagawa}}\ and\ \bibinfo
  {author} {\bibfnamefont{M.}~\bibnamefont{Ueda}},\ }%
  \bibfield{journal}{%
  \Doi{10.1103/PhysRevA.47.5138}{\bibinfo {journal} {Phys. Rev. A}}\ }%
  \textbf{\bibinfo {volume} {47}},\ \bibinfo {pages} {5138} (\bibinfo {year}
  {1993})%
  \bibAnnoteFile{NoStop}{kitagawa1993squeezed}%
\bibitem{wineland1994squeezed}%
  \BibitemOpen
  \bibfield{author}{%
  \bibinfo {author} {\bibfnamefont{D.}~\bibnamefont{Wineland}}, \bibinfo
  {author} {\bibfnamefont{J.}~\bibnamefont{Bollinger}}, \bibinfo {author}
  {\bibfnamefont{W.}~\bibnamefont{Itano}},\ and\ \bibinfo {author}
  {\bibfnamefont{D.}~\bibnamefont{Heinzen}},\ }%
  \bibfield{journal}{%
  \Doi{10.1103/PhysRevA.50.67}{\bibinfo {journal} {Phys. Rev. A}}\ }%
  \textbf{\bibinfo {volume} {50}},\ \bibinfo {pages} {67} (\bibinfo {year}
  {1994})%
  \bibAnnoteFile{NoStop}{wineland1994squeezed}%
\bibitem{sorensen2001many}%
  \BibitemOpen
  \bibfield{author}{%
  \bibinfo {author} {\bibfnamefont{A.}~\bibnamefont{S{\o}rensen}}, \bibinfo
  {author} {\bibfnamefont{L.-M.}\ \bibnamefont{Duan}}, \bibinfo {author}
  {\bibfnamefont{J.}~\bibnamefont{Cirac}},\ and\ \bibinfo {author}
  {\bibfnamefont{P.}~\bibnamefont{Zoller}},\ }%
  \bibfield{journal}{%
  \Doi{10.1038/35051038}{\bibinfo {journal} {Nature}}\ }%
  \textbf{\bibinfo {volume} {409}},\ \bibinfo {pages} {63} (\bibinfo {year}
  {2001})%
  \bibAnnoteFile{NoStop}{sorensen2001many}%
\bibitem{esteve2008squeezing}%
  \BibitemOpen
  \bibfield{author}{%
  \bibinfo {author} {\bibfnamefont{J.}~\bibnamefont{Esteve}}, \bibinfo {author}
  {\bibfnamefont{C.}~\bibnamefont{Gross}}, \bibinfo {author}
  {\bibfnamefont{A.}~\bibnamefont{Weller}}, \bibinfo {author}
  {\bibfnamefont{S.}~\bibnamefont{Giovanazzi}},\ and\ \bibinfo {author}
  {\bibfnamefont{M.}~\bibnamefont{Oberthaler}},\ }%
  \bibfield{journal}{%
  \Doi{10.1038/nature07332}{\bibinfo {journal} {Nature}}\ }%
  \textbf{\bibinfo {volume} {455}},\ \bibinfo {pages} {1216} (\bibinfo {year}
  {2008})%
  \bibAnnoteFile{NoStop}{esteve2008squeezing}%
\bibitem{appel2009mesoscopic}%
  \BibitemOpen
  \bibfield{author}{%
  \bibinfo {author} {\bibfnamefont{J.}~\bibnamefont{Appel}}, \bibinfo {author}
  {\bibfnamefont{P.~J.}\ \bibnamefont{Windpassinger}}, \bibinfo {author}
  {\bibfnamefont{D.}~\bibnamefont{Oblak}}, \bibinfo {author}
  {\bibfnamefont{U.~B.}\ \bibnamefont{Hoff}}, \bibinfo {author}
  {\bibfnamefont{N.}~\bibnamefont{Kj{\ae}rgaard}},\ and\ \bibinfo {author}
  {\bibfnamefont{E.~S.}\ \bibnamefont{Polzik}},\ }%
  \bibfield{journal}{%
  \bibinfo {journal} {Proc. Natl. Acad. Sci. USA}\ }%
  \textbf{\bibinfo {volume} {106}},\ \bibinfo {pages} {10960} (\bibinfo {year}
  {2009})%
  \bibAnnoteFile{NoStop}{appel2009mesoscopic}%
\bibitem{gross2010nonlinear}%
  \BibitemOpen
  \bibfield{author}{%
  \bibinfo {author} {\bibfnamefont{C.}~\bibnamefont{Gross}}, \bibinfo {author}
  {\bibfnamefont{T.}~\bibnamefont{Zibold}}, \bibinfo {author}
  {\bibfnamefont{E.}~\bibnamefont{Nicklas}}, \bibinfo {author}
  {\bibfnamefont{J.}~\bibnamefont{Esteve}},\ and\ \bibinfo {author}
  {\bibfnamefont{M.~K.}\ \bibnamefont{Oberthaler}},\ }%
  \bibfield{journal}{%
  \Doi{10.1038/nature08919}{\bibinfo {journal} {Nature}}\ }%
  \textbf{\bibinfo {volume} {464}},\ \bibinfo {pages} {1165} (\bibinfo {year}
  {2010})%
  \bibAnnoteFile{NoStop}{gross2010nonlinear}%
\bibitem{riedel2010atom}%
  \BibitemOpen
  \bibfield{author}{%
  \bibinfo {author} {\bibfnamefont{M.~F.}\ \bibnamefont{Riedel}}, \bibinfo
  {author} {\bibfnamefont{P.}~\bibnamefont{B{\"o}hi}}, \bibinfo {author}
  {\bibfnamefont{Y.}~\bibnamefont{Li}}, \bibinfo {author}
  {\bibfnamefont{T.~W.}\ \bibnamefont{H{\"a}nsch}}, \bibinfo {author}
  {\bibfnamefont{A.}~\bibnamefont{Sinatra}},\ and\ \bibinfo {author}
  {\bibfnamefont{P.}~\bibnamefont{Treutlein}},\ }%
  \bibfield{journal}{%
  \Doi{10.1038/nature08988}{\bibinfo {journal} {Nature}}\ }%
  \textbf{\bibinfo {volume} {464}},\ \bibinfo {pages} {1170} (\bibinfo {year}
  {2010})%
  \bibAnnoteFile{NoStop}{riedel2010atom}%
\bibitem{oat1}%
  \BibitemOpen
  \bibfield{author}{%
  \bibinfo {author} {\bibfnamefont{M.}~\bibnamefont{Takeuchi}}, \bibinfo
  {author} {\bibfnamefont{S.}~\bibnamefont{Ichihara}}, \bibinfo {author}
  {\bibfnamefont{T.}~\bibnamefont{Takano}}, \bibinfo {author}
  {\bibfnamefont{M.}~\bibnamefont{Kumakura}}, \bibinfo {author}
  {\bibfnamefont{T.}~\bibnamefont{Yabuzaki}},\ and\ \bibinfo {author}
  {\bibfnamefont{Y.}~\bibnamefont{Takahashi}},\ }%
  \bibfield{journal}{%
  \bibinfo {journal} {Phys. Rev. Lett.}\ }%
  \textbf{\bibinfo {volume} {94}},\ \bibinfo {pages} {023003} (\bibinfo {year}
  {2005})%
  \bibAnnoteFile{NoStop}{oat1}%
\bibitem{nori}%
  \BibitemOpen
  \bibfield{author}{%
  \bibinfo {author} {\bibfnamefont{J.}~\bibnamefont{Ma}}, \bibinfo {author}
  {\bibfnamefont{X.}~\bibnamefont{Wang}}, \bibinfo {author}
  {\bibfnamefont{C.}~\bibnamefont{Sun}},\ and\ \bibinfo {author}
  {\bibfnamefont{F.}~\bibnamefont{Nori}},\ }%
  \bibfield{journal}{%
  \Doi{http://dx.doi.org/10.1016/j.physrep.2011.08.003}{\bibinfo {journal}
  {Physics Reports}}\ }%
  \textbf{\bibinfo {volume} {509}},\ \bibinfo {pages} {89 } (\bibinfo {year}
  {2011}),\ ISSN \bibinfo {issn} {0370-1573}%
  \bibAnnoteFile{NoStop}{nori}%
\bibitem{polzik}%
  \BibitemOpen
  \bibfield{author}{%
  \bibinfo {author} {\bibfnamefont{T.}~\bibnamefont{Fernholz}}, \bibinfo
  {author} {\bibfnamefont{H.}~\bibnamefont{Krauter}}, \bibinfo {author}
  {\bibfnamefont{K.}~\bibnamefont{Jensen}}, \bibinfo {author}
  {\bibfnamefont{J.~F.}\ \bibnamefont{Sherson}}, \bibinfo {author}
  {\bibfnamefont{A.~S.}\ \bibnamefont{S\o{}rensen}},\ and\ \bibinfo {author}
  {\bibfnamefont{E.~S.}\ \bibnamefont{Polzik}},\ }%
  \bibfield{journal}{%
  \Doi{10.1103/PhysRevLett.101.073601}{\bibinfo {journal} {Phys. Rev. Lett.}}\
  }%
  \textbf{\bibinfo {volume} {101}},\ \bibinfo {pages} {073601} (\bibinfo {year}
  {2008})%
  \bibAnnoteFile{NoStop}{polzik}%
\bibitem{radcliffe1971some}%
  \BibitemOpen
  \bibfield{author}{%
  \bibinfo {author} {\bibfnamefont{J.}~\bibnamefont{Radcliffe}},\ }%
  \bibfield{journal}{%
  \Doi{10.1088/0305-4470/4/3/009}{\bibinfo {journal} {Journal of Physics A:
  General Physics}}\ }%
  \textbf{\bibinfo {volume} {4}},\ \bibinfo {pages} {313} (\bibinfo {year}
  {1971})%
  \bibAnnoteFile{NoStop}{radcliffe1971some}%
\bibitem{pezze2009entanglement}%
  \BibitemOpen
  \bibfield{author}{%
  \bibinfo {author} {\bibfnamefont{L.}~\bibnamefont{Pezz{\'e}}}\ and\ \bibinfo
  {author} {\bibfnamefont{A.}~\bibnamefont{Smerzi}},\ }%
  \bibfield{journal}{%
  \Doi{10.1103/PhysRevLett.102.100401}{\bibinfo {journal} {Phys. Rev. Lett.}}\
  }%
  \textbf{\bibinfo {volume} {102}},\ \bibinfo {pages} {100401} (\bibinfo {year}
  {2009})%
  \bibAnnoteFile{NoStop}{pezze2009entanglement}%
\bibitem{greenberger1989going}%
  \BibitemOpen
  \bibfield{author}{%
  \bibinfo {author} {\bibfnamefont{D.~M.}\ \bibnamefont{Greenberger}}, \bibinfo
  {author} {\bibfnamefont{M.~A.}\ \bibnamefont{Horne}},\ and\ \bibinfo {author}
  {\bibfnamefont{A.}~\bibnamefont{Zeilinger}},\ }%
  \emph{\bibinfo {title} {Going beyond Bell’s theorem}}\ (\bibinfo
  {publisher} {Springer},\ \bibinfo {year} {1989})\ pp.\ \bibinfo {pages}
  {69--72}%
  \bibAnnoteFile{NoStop}{greenberger1989going}%
\bibitem{smerzi_ob}%
  \BibitemOpen
  \bibfield{author}{%
  \bibinfo {author} {\bibfnamefont{H.}~\bibnamefont{Strobel}}, \bibinfo
  {author} {\bibfnamefont{W.}~\bibnamefont{Muessel}}, \bibinfo {author}
  {\bibfnamefont{D.}~\bibnamefont{Linnemann}}, \bibinfo {author}
  {\bibfnamefont{T.}~\bibnamefont{Zibold}}, \bibinfo {author}
  {\bibfnamefont{D.~B.}\ \bibnamefont{Hume}}, \bibinfo {author}
  {\bibfnamefont{L.}~\bibnamefont{Pezz\'e}}, \bibinfo {author}
  {\bibfnamefont{A.}~\bibnamefont{Smerzi}},\ and\ \bibinfo {author}
  {\bibfnamefont{M.~K.}\ \bibnamefont{Oberthaler}},\ }%
  \bibfield{journal}{%
  \Doi{10.1126/science.1250147}{\bibinfo {journal} {Science}}\ }%
  \textbf{\bibinfo {volume} {345}},\ \bibinfo {pages} {424} (\bibinfo {year}
  {2014})%
  \bibAnnoteFile{NoStop}{smerzi_ob}%
\bibitem{schmied_rmp}%
  \BibitemOpen
  \bibfield{author}{%
  \bibinfo {author} {\bibfnamefont{A.~D.}\ \bibnamefont{Cronin}}, \bibinfo
  {author} {\bibfnamefont{J.}~\bibnamefont{Schmiedmayer}},\ and\ \bibinfo
  {author} {\bibfnamefont{D.~E.}\ \bibnamefont{Pritchard}},\ }%
  \bibfield{journal}{%
  \bibinfo {journal} {Rev. Mod. Phys.}\ }%
  \textbf{\bibinfo {volume} {81}},\ \bibinfo {pages} {1051} (\bibinfo {year}
  {2009})%
  \bibAnnoteFile{NoStop}{schmied_rmp}%
\bibitem{pdc1}%
  \BibitemOpen
  \bibfield{author}{%
  \bibinfo {author} {\bibfnamefont{D.~C.}\ \bibnamefont{Burnham}}\ and\
  \bibinfo {author} {\bibfnamefont{D.~L.}\ \bibnamefont{Weinberg}},\ }%
  \bibfield{journal}{%
  \bibinfo {journal} {Phys. Rev. Lett.}\ }%
  \textbf{\bibinfo {volume} {25}},\ \bibinfo {pages} {84} (\bibinfo {year}
  {1970})%
  \bibAnnoteFile{NoStop}{pdc1}%
\bibitem{pdc2}%
  \BibitemOpen
  \bibfield{author}{%
  \bibinfo {author} {\bibfnamefont{P.~G.}\ \bibnamefont{Kwiat}}, \bibinfo
  {author} {\bibfnamefont{K.}~\bibnamefont{Mattle}}, \bibinfo {author}
  {\bibfnamefont{H.}~\bibnamefont{Weinfurter}}, \bibinfo {author}
  {\bibfnamefont{A.}~\bibnamefont{Zeilinger}}, \bibinfo {author}
  {\bibfnamefont{A.~V.}\ \bibnamefont{Sergienko}},\ and\ \bibinfo {author}
  {\bibfnamefont{Y.}~\bibnamefont{Shih}},\ }%
  \bibfield{journal}{%
  \bibinfo {journal} {Phys. Rev. Lett.}\ }%
  \textbf{\bibinfo {volume} {75}},\ \bibinfo {pages} {4337} (\bibinfo {year}
  {1995})%
  \bibAnnoteFile{NoStop}{pdc2}%
\bibitem{berrada2013integrated}%
  \BibitemOpen
  \bibfield{author}{%
  \bibinfo {author} {\bibfnamefont{T.}~\bibnamefont{Berrada}}, \bibinfo
  {author} {\bibfnamefont{S.}~\bibnamefont{van Frank}}, \bibinfo {author}
  {\bibfnamefont{R.}~\bibnamefont{B{\"u}cker}}, \bibinfo {author}
  {\bibfnamefont{T.}~\bibnamefont{Schumm}}, \bibinfo {author}
  {\bibfnamefont{J.-F.}\ \bibnamefont{Schaff}},\ and\ \bibinfo {author}
  {\bibfnamefont{J.}~\bibnamefont{Schmiedmayer}},\ }%
  \bibfield{journal}{%
  \bibinfo {journal} {Nat. Commun.}\ }%
  \textbf{\bibinfo {volume} {4}} (\bibinfo {year} {2013})%
  \bibAnnoteFile{NoStop}{berrada2013integrated}%
\bibitem{smerzi_twin}%
  \BibitemOpen
  \bibfield{author}{%
  \bibinfo {author} {\bibfnamefont{B.}~\bibnamefont{L\"ucke}}, \bibinfo
  {author} {\bibfnamefont{M.}~\bibnamefont{Scherer}}, \bibinfo {author}
  {\bibfnamefont{J.}~\bibnamefont{Kruse}}, \bibinfo {author}
  {\bibfnamefont{L.}~\bibnamefont{Pezz\'e}}, \bibinfo {author}
  {\bibfnamefont{F.}~\bibnamefont{Deuretzbacher}}, \bibinfo {author}
  {\bibfnamefont{P.}~\bibnamefont{Hyllus}}, \bibinfo {author}
  {\bibfnamefont{O.}~\bibnamefont{Topic}}, \bibinfo {author}
  {\bibfnamefont{J.}~\bibnamefont{Peise}}, \bibinfo {author}
  {\bibfnamefont{W.}~\bibnamefont{Ertmer}}, \bibinfo {author}
  {\bibfnamefont{J.}~\bibnamefont{Arlt}}, \bibinfo {author}
  {\bibfnamefont{L.}~\bibnamefont{Santos}}, \bibinfo {author}
  {\bibfnamefont{A.}~\bibnamefont{Smerzi}},\ and\ \bibinfo {author}
  {\bibfnamefont{C.}~\bibnamefont{Klempt}},\ }%
  \bibfield{journal}{%
  \bibinfo {journal} {Science}\ }%
  \textbf{\bibinfo {volume} {334}},\ \bibinfo {pages} {773} (\bibinfo {year}
  {2011})%
  \bibAnnoteFile{NoStop}{smerzi_twin}%
\bibitem{rotator}%
  \BibitemOpen
  \bibfield{author}{%
  \bibinfo {author} {\bibfnamefont{B.~C.}\ \bibnamefont{Sanders}},\ }%
  \bibfield{journal}{%
  \bibinfo {journal} {Phys. Rev. A}\ }%
  \textbf{\bibinfo {volume} {40}},\ \bibinfo {pages} {2417} (\bibinfo {year}
  {1989})%
  \bibAnnoteFile{NoStop}{rotator}%
\bibitem{non_condensed}%
  \BibitemOpen
  \bibfield{author}{%
  \bibinfo {author} {\bibfnamefont{T.}~\bibnamefont{Takano}}, \bibinfo {author}
  {\bibfnamefont{M.}~\bibnamefont{Fuyama}}, \bibinfo {author}
  {\bibfnamefont{R.}~\bibnamefont{Namiki}},\ and\ \bibinfo {author}
  {\bibfnamefont{Y.}~\bibnamefont{Takahashi}},\ }%
  \bibfield{journal}{%
  \bibinfo {journal} {Phys. Rev. Lett.}\ }%
  \textbf{\bibinfo {volume} {102}},\ \bibinfo {pages} {033601} (\bibinfo {year}
  {2009})%
  \bibAnnoteFile{NoStop}{non_condensed}%
\bibitem{kolo}%
  \BibitemOpen
  \bibfield{author}{%
  \bibinfo {author} {\bibfnamefont{J.~B.}\ \bibnamefont{Brask}}, \bibinfo
  {author} {\bibfnamefont{R.}~\bibnamefont{Chaves}},\ and\ \bibinfo {author}
  {\bibfnamefont{J.}~\bibnamefont{Ko{\l}ody\'nski}},\ }%
  \bibinfo {journal} {arXiv:1411.0716}%
  \bibAnnoteFile{NoStop}{kolo}%
\bibitem{ferrini}%
  \BibitemOpen
\bibfield{journal}{%
    }%
  \bibfield{author}{%
  \bibinfo {author} {\bibfnamefont{K.}~\bibnamefont{Paw{\l}owski}}, \bibinfo
  {author} {\bibfnamefont{D.}~\bibnamefont{Spehner}}, \bibinfo {author}
  {\bibfnamefont{A.}~\bibnamefont{Minguzzi}},\ and\ \bibinfo {author}
  {\bibfnamefont{G.}~\bibnamefont{Ferrini}},\ }%
  \bibfield{journal}{%
  \bibinfo {journal} {Phys. Rev. A}\ }%
  \textbf{\bibinfo {volume} {88}},\ \bibinfo {pages} {013606} (\bibinfo {year}
  {2013})%
  \bibAnnoteFile{NoStop}{ferrini}%
\bibitem{giovannetti2006quantum}%
  \BibitemOpen
  \bibfield{author}{%
  \bibinfo {author} {\bibfnamefont{V.}~\bibnamefont{Giovannetti}}, \bibinfo
  {author} {\bibfnamefont{S.}~\bibnamefont{Lloyd}},\ and\ \bibinfo {author}
  {\bibfnamefont{L.}~\bibnamefont{Maccone}},\ }%
  \bibfield{journal}{%
  \Doi{10.1103/PhysRevLett.96.010401}{\bibinfo {journal} {Phys. Rev. Lett.}}\
  }%
  \textbf{\bibinfo {volume} {96}},\ \bibinfo {pages} {010401} (\bibinfo {year}
  {2006})%
  \bibAnnoteFile{NoStop}{giovannetti2006quantum}%
\bibitem{Note1}%
  \BibitemOpen
  \bibinfo {note} {Indeed, the variation of the phase $\varphi (m)=\alpha m^2$
  can be estimated by its derivative $\protect \frac {\partial \varphi
  }{\partial m}\lesssim 2\pi $. For the initial coherent state, the maximal
  $m\sim N^{\protect \frac 12}$, giving $\alpha \lesssim N^{-\protect \frac
  12}$}%
  \bibAnnoteFile{NoStop}{Note1}%
\bibitem{holevo2011probabilistic}%
  \BibitemOpen
  \bibfield{author}{%
  \bibinfo {author} {\bibfnamefont{A.}~\bibnamefont{Holevo}},\ }%
  \emph{\bibinfo {title} {Probabilistic and Statistical Aspects of Quantum
  Theory}}\ (\bibinfo {publisher} {Publications of Scuola Normale Superiore},\
  \bibinfo {year} {2011})%
  \bibAnnoteFile{NoStop}{holevo2011probabilistic}%
\bibitem{braunstein1994statistical}%
  \BibitemOpen
  \bibfield{author}{%
  \bibinfo {author} {\bibfnamefont{S.~L.}\ \bibnamefont{Braunstein}}\ and\
  \bibinfo {author} {\bibfnamefont{C.~M.}\ \bibnamefont{Caves}},\ }%
  \bibfield{journal}{%
  \Doi{10.1103/PhysRevLett.72.3439}{\bibinfo {journal} {Phys. Rev. Lett.}}\ }%
  \textbf{\bibinfo {volume} {72}},\ \bibinfo {pages} {3439} (\bibinfo {year}
  {1994})%
  \bibAnnoteFile{NoStop}{braunstein1994statistical}%
\bibitem{Note2}%
  \BibitemOpen
  \bibinfo {note} {We have used the fact that the $C_m$'s and the imprinted
  phase are symmetric with respect to $m=0$.}%
  \bibAnnoteFile{Stop}{Note2}%
\bibitem{fesch_jul}%
  \BibitemOpen
  \bibfield{author}{%
  \bibinfo {author} {\bibfnamefont{C.}~\bibnamefont{Chin}}, \bibinfo {author}
  {\bibfnamefont{R.}~\bibnamefont{Grimm}}, \bibinfo {author}
  {\bibfnamefont{P.}~\bibnamefont{Julienne}},\ and\ \bibinfo {author}
  {\bibfnamefont{E.}~\bibnamefont{Tiesinga}},\ }%
  \bibfield{journal}{%
  \bibinfo {journal} {Rev. Mod. Phys.}\ }%
  \textbf{\bibinfo {volume} {82}},\ \bibinfo {pages} {1225} (\bibinfo {year}
  {2010})%
  \bibAnnoteFile{NoStop}{fesch_jul}%
\bibitem{dalfovo}%
  \BibitemOpen
  \bibfield{author}{%
  \bibinfo {author} {\bibfnamefont{F.}~\bibnamefont{Dalfovo}}, \bibinfo
  {author} {\bibfnamefont{S.}~\bibnamefont{Giorgini}}, \bibinfo {author}
  {\bibfnamefont{L.~P.}\ \bibnamefont{Pitaevskii}},\ and\ \bibinfo {author}
  {\bibfnamefont{S.}~\bibnamefont{Stringari}},\ }%
  \bibfield{journal}{%
  \bibinfo {journal} {Rev. Mod. Phys.}\ }%
  \textbf{\bibinfo {volume} {71}},\ \bibinfo {pages} {463} (\bibinfo {year}
  {1999})%
  \bibAnnoteFile{NoStop}{dalfovo}%
\bibitem{zwerger}%
  \BibitemOpen
  \bibfield{author}{%
  \bibinfo {author} {\bibfnamefont{I.}~\bibnamefont{Bloch}}, \bibinfo {author}
  {\bibfnamefont{J.}~\bibnamefont{Dalibard}},\ and\ \bibinfo {author}
  {\bibfnamefont{W.}~\bibnamefont{Zwerger}},\ }%
  \bibfield{journal}{%
  \bibinfo {journal} {Rev. Mod. Phys.}\ }%
  \textbf{\bibinfo {volume} {80}},\ \bibinfo {pages} {885} (\bibinfo {year}
  {2008})%
  \bibAnnoteFile{NoStop}{zwerger}%
\bibitem{szan}%
  \BibitemOpen
  \bibfield{author}{%
  \bibinfo {author} {\bibfnamefont{P.}~\bibnamefont{Sza\ifmmode~\acute{n}\else
  \'{n}\fi{}kowski}}, \bibinfo {author}
  {\bibfnamefont{M.}~\bibnamefont{Trippenbach}},\ and\ \bibinfo {author}
  {\bibfnamefont{J.}~\bibnamefont{Chwede\ifmmode~\acute{n}\else
  \'{n}\fi{}czuk}},\ }%
  \bibfield{journal}{%
  \bibinfo {journal} {Phys. Rev. A}\ }%
  \textbf{\bibinfo {volume} {90}},\ \bibinfo {pages} {063619} (\bibinfo {year}
  {2014})%
  \bibAnnoteFile{NoStop}{szan}%
\bibitem{obert_detect}%
  \BibitemOpen
  \bibfield{author}{%
  \bibinfo {author} {\bibfnamefont{D.~B.}\ \bibnamefont{Hume}}, \bibinfo
  {author} {\bibfnamefont{I.}~\bibnamefont{Stroescu}}, \bibinfo {author}
  {\bibfnamefont{M.}~\bibnamefont{Joos}}, \bibinfo {author}
  {\bibfnamefont{W.}~\bibnamefont{Muessel}}, \bibinfo {author}
  {\bibfnamefont{H.}~\bibnamefont{Strobel}},\ and\ \bibinfo {author}
  {\bibfnamefont{M.~K.}\ \bibnamefont{Oberthaler}},\ }%
  \bibfield{journal}{%
  \bibinfo {journal} {Phys. Rev. Lett.}\ }%
  \textbf{\bibinfo {volume} {111}},\ \bibinfo {pages} {253001} (\bibinfo {year}
  {2013})%
  \bibAnnoteFile{NoStop}{obert_detect}%
\bibitem{sheet}%
  \BibitemOpen
  \bibfield{author}{%
  \bibinfo {author} {\bibfnamefont{R.}~\bibnamefont{B\"ucker}}, \bibinfo
  {author} {\bibfnamefont{A.}~\bibnamefont{Perrin}}, \bibinfo {author}
  {\bibfnamefont{S.}~\bibnamefont{Manz}}, \bibinfo {author}
  {\bibfnamefont{T.}~\bibnamefont{Betz}}, \bibinfo {author}
  {\bibfnamefont{C.}~\bibnamefont{Koller}}, \bibinfo {author}
  {\bibfnamefont{T.}~\bibnamefont{Plisson}}, \bibinfo {author}
  {\bibfnamefont{J.}~\bibnamefont{Rottmann}}, \bibinfo {author}
  {\bibfnamefont{T.}~\bibnamefont{Schumm}},\ and\ \bibinfo {author}
  {\bibfnamefont{J.}~\bibnamefont{Schmiedmayer}},\ }%
  \bibfield{journal}{%
  \bibinfo {journal} {New J. Phys.}\ }%
  \textbf{\bibinfo {volume} {11}},\ \bibinfo {pages} {103039} (\bibinfo {year}
  {2009})%
  \bibAnnoteFile{NoStop}{sheet}%
\bibitem{bloch}%
  \BibitemOpen
  \bibfield{author}{%
  \bibinfo {author} {\bibfnamefont{J.~F.}\ \bibnamefont{Sherson}}, \bibinfo
  {author} {\bibfnamefont{C.}~\bibnamefont{Weitenberg}}, \bibinfo {author}
  {\bibfnamefont{M.}~\bibnamefont{Endres}}, \bibinfo {author}
  {\bibfnamefont{M.}~\bibnamefont{Cheneau}}, \bibinfo {author}
  {\bibfnamefont{I.}~\bibnamefont{Bloch}},\ and\ \bibinfo {author}
  {\bibfnamefont{S.}~\bibnamefont{Kuhr}},\ }%
  \bibfield{journal}{%
  \bibinfo {journal} {Nature}\ }%
  \textbf{\bibinfo {volume} {467}},\ \bibinfo {pages} {68} (\bibinfo {year}
  {2010})%
  \bibAnnoteFile{NoStop}{bloch}%
\bibitem{paris_han}%
  \BibitemOpen
  \bibfield{author}{%
  \bibinfo {author} {\bibfnamefont{M.}~\bibnamefont{Schellekens}}, \bibinfo
  {author} {\bibfnamefont{R.}~\bibnamefont{Hoppeler}}, \bibinfo {author}
  {\bibfnamefont{A.}~\bibnamefont{Perrin}}, \bibinfo {author}
  {\bibfnamefont{J.~V.}\ \bibnamefont{Gomes}}, \bibinfo {author}
  {\bibfnamefont{D.}~\bibnamefont{Boiron}}, \bibinfo {author}
  {\bibfnamefont{A.}~\bibnamefont{Aspect}},\ and\ \bibinfo {author}
  {\bibfnamefont{C.~I.}\ \bibnamefont{Westbrook}},\ }%
  \bibfield{journal}{%
  \bibinfo {journal} {Science}\ }%
  \textbf{\bibinfo {volume} {310}},\ \bibinfo {pages} {648} (\bibinfo {year}
  {2005})%
  \bibAnnoteFile{NoStop}{paris_han}%
\bibitem{wasak2015interferometry}%
  \BibitemOpen
  \bibfield{author}{%
  \bibinfo {author} {\bibfnamefont{T.}~\bibnamefont{Wasak}}, \bibinfo {author}
  {\bibfnamefont{P.}~\bibnamefont{Sza{\'n}kowski}},\ and\ \bibinfo {author}
  {\bibfnamefont{J.}~\bibnamefont{Chwede{\'n}czuk}},\ }%
  \bibfield{journal}{%
  \bibinfo {journal} {Phys. Rev. A}\ }%
  \textbf{\bibinfo {volume} {91}},\ \bibinfo {pages} {043619} (\bibinfo {year}
  {2015})%
  \bibAnnoteFile{NoStop}{wasak2015interferometry}%
\bibitem{Note3}%
  \BibitemOpen
  \bibinfo {note} {The fidelity between two probability distributions $p_n$ and
  $q_n$ is defined as $f=\DOTSB \sum@ \slimits@ _n\protect \sqrt {p_nq_n}$.}%
  \bibAnnoteFile{Stop}{Note3}%
\end{thebibliography}
\end{document}